\documentclass[preprint,showpacs,nofootinbib,12pt]{iopart}
\usepackage{iopams,setstack}
\usepackage{xcolor}
\usepackage[hypertex]{hyperref}
\usepackage{graphicx,subfigure}
\usepackage{multicol}
\newcommand{\m}{\mathbf}

\begin{document}

\title[]{The vortex pinning on the cylindrical defects and the electronic structure of the vortex core}

\author{V.~L. Kulinskii, D.~Yu. Panchenko}
\address{Department of Theoretical Physics, Odessa National University, Dvoryanskaya 2, 65026 Odessa, Ukraine}
\eads{kulinskij@onu.edu.ua, dpanchenko@onu.edu.ua}

\begin{abstract}
The model of the Abrikosov vortex pinning on a cylindrical defect is proposed. It is shown that in the limit $\varkappa \gg 1$ the potential for the vortex core excitations can be treated in terms of the zero-range potentials method. Using the variational method the estimates for the energy of pinning, the pinning force and the density of critical current defect are obtained.
\end{abstract}
\submitto{J. Phys. A: Math. Theor.}


\section{Introduction}
The pinning of Abrikosov vortices on the defects influences the static and dynamic properties  of the superconductors. In particular the critical current density $j_c$ which determines the stability of the superconducting phase depends on the pinning. In order to get the superconductor with higher value of $j_c$ the defects of the crystal structure should be created since they are the pinning centers. If there are too much defects the superconductivity is destroyed \cite{book_abrikosov,book_pinning_matsushita}. Evidently the pinning essentially depends on type of defects interacting with vortex. In this connection the pinning can be of the following types:
 single -- vortex pinning on the microscopical (nanometer scale) inclusion of normal phase \cite{sc_strongpinning_iop2008};
superficial pinning on a film inhomogeneity \cite{sc_geometricsurfacepinning_jetp2003};
pinning on a lengthy flat defects of twin boundary type \cite{sc_epitaxial_prl1990} or antiphase boundary defect \cite{sc_antiphase_prl2000};
pinning on a linear (quasionedimensional) defects, in particular on screw and edge dislocation \cite{sc_strongpinning_prl2000, sc_lineardefects_prl2001, sc_epitaxialfilms_lowtemp2002};
The most effective pining occurs when the geometry of the defect coincides  with the geometry of the vortex, i.e. for the columnar defect parallel to the axis of the vortex \cite{sc_columndefection_prl1991}. Due to progress in nanotechnology now it is possible to implant the material with the defects of controllable characteristics of atomic size scale. For the physical situation considered the most interesting is the perforating the material with the cylindrical pores of radius $R$ via the bombardment by fast ions or highly energetic protons \cite{sc_columndefectproton_prl1990,sc_columndefection_prl1991}.

The phenomenological approach to estimate the force (per unit length) and the corresponding density of critical current in the case when the radius of the cylinder $R\ll \lambda$ was proposed in  \cite{sc_vortexpincavity_schmidt_jetp1971} and extended to the case of arbitrary large values of $R$ in \cite{sc_vortexpinning_prb2000}. These approaches were based on the phenomenological Landau-Ginzburg theory and therefore use the assumptions about the order parameter $\Delta(\m{r})$, which is the main characteristic of electronic structure of the vortex. The case of $R\lesssim \xi$ was considered within the quasiclassical approximation  for the Bogolyubov-de Gennes (BdG) theory \cite{sc_vortexpinning_prb2009}. The characteristic scale of the spatial variation of $\Delta(\m{r})$ is the correlation length $\xi$. Although in many works the specific spatial dependence of $\Delta(\m{r})$ is not of principal interest the consideration of the transition to the limit $\varkappa \to \infty$ where two characteristic scales $R$ and $\xi$ merge does depend on the assumption of the spatial distribution of the order parameter $\Delta$. According to \cite{sc_vortexcorevolovik_jetplet1993} the structure of $\Delta(\m{r})$ can be even more complex and is characterized by additional scale $\xi_1<\xi$, which separates the regions at the point where the jump of the derivative of the order parameter occurs. The quantity $\xi_1$ also determines the distance where the supercurrent density reaches its maximum \cite{sc_vortexmuspectrores_rmp2000,sc_vorticestruct_jphyscondmat2004}. So we treat the distance $\xi_1$ as the characteristic length scale of the vortex core.
In its turn the increase of the slope of $\Delta(\m{r})$ in the region $r<\xi_1$ is reflected in ``Kramer-Pesch effect`` \cite{sc_vortexcorekramerpesch_zphys1974}. It diverges in the quantum limit $T\to 0$ where $k^{-1}_{F}\simeq\xi_{1}\ll\xi_{BCS}$ because of the oscillations of the Bogolyubov wave functions \cite{sc_vortexcorevolovik_jetplet1993}. The shrinking of the core region leads
to a reduction in the number of bound states \cite{sc_vortexcorelowexcit_prl1998}.

Thus the structure of the vortex core is far from trivial and should be treated correctly, especially in the limit $\varkappa \to \infty$ due to point-like singularities in the spatial distribution of physical quantities.

As the vortex core plays the role of the defect the appearance of the bound states is expected in analogy with the localized states near the defects in semiconductors \cite{book_semicondanselm}.
There exists the branch of the localized excitations \cite{sc_carollidegennes_physlet1964} which can be called ``magnetic``. Another branch of bound states is possible due to the repulsive interaction with the defect in the vortex core \cite{sc_vortexcorespectr_physb2002}.

The aim of this paper is to formulate the model for pinning which
treats the appearance of the bound states in the limit $\varkappa \to \infty$. This limiting case can be considered within the zero range potential method \cite{book_bazeldovichperelomov_en}. We also neglect the temperature effect as we work in the quantum limit \cite{sc_vortexmuspectrores_rmp2000}. Here only lowest bound states are populated and the behavior of the slope of $\Delta(r)$ at the vortex center becomes very steep.
The structure of the paper is as follows.
In Section~\ref{sec_ham} we consider the relation between the BdG Hamiltonian  for the qusiparticle excitations and the Aharonov-Bohm (AB) Hamiltonian  for the particle in the field of the localized magnetic flux. We show that these hamiltonians are equivalent for the low lying energy states localized near the vortex core where the contribution due to order parameter $\Delta$ can be neglected. We apply the treatment of the AB Hamiltonian developed in \cite{abeffect_deltainterct_jmathphys1998} for the investigation of these bound states. Basing on the physical nature these states we differentiate between the boundary conditions corresponding to the AB effect and to the case of the Abrikosov vortex (Ab-vortex). In Section~\ref{sec_bound} two types of the bound states are considered and their susceptibility to the magnetic flux is analyzed. The results are summarized  in the concluding section.

\section{The Hamiltonian for the bounded states in the vicinity of localized vortex}\label{sec_ham}
The pinning mechanism depends on the electronic structure of the Ab-vortex core, the type of a defect and the interaction between them. In  simple situation such interaction is due to that between electrons of the vortex core and the defect. Specific form of the interaction potential is determined by the interaction between the excitations which form the normal Ab-vortex core and the condensate of the Cooper pairs.

The Bogolyubov-de Gennes Hamiltonian (BdGH), which describes the energy levels for such excitations is \cite{sc_carollidegennes_physlet1964}:
\begin{equation}\label{ham0}
\hat{H}=\sigma_{z}\left\{\frac{1}{2m}\left(\,\hat{\mathbf{p}}-
\sigma_{z}\frac{e}{c}\mathbf{A}-
\frac{1}{2}\sigma_{z}\hbar\nabla\theta\,\right)^{2}-E_{F}\right\}+\sigma_{x}\Delta(r)
\end{equation}
where $E_F$ is the Fermi energy, the vector potential $\m{A}$ describes the applied magnetic field of order $H_{c_{1}}$ while the gradient term is for the magnetic field localized in the vortex \cite{sc_schmidt_supercond}.

Since \[A\sim H_{c_{1}}r;\quad eA/c\hbar\nabla\theta\sim H\xi^{2}/\Phi_{0}\sim H/H_{c_{2}}\ll 1\,,\] then we can neglect $A$ and rewrite the BdGH \eref{ham0} in the form (see \cite{sc_carollidegennes_physlet1964})
\begin{equation}\label{radham}
\hat{H}=\sigma_{z}\frac{\hbar^{2}}{2m}\Bigl\{-\frac{d^{2}}{dr^{2}}-\frac{1}{r}\frac{d}{dr}
+\left(\mu-\frac{1}{2}\sigma_{z}\right)^{2}\frac{1}{r^{2}}-E_{F}\Bigr\}+\sigma_{x}\Delta(r)\,,
\end{equation}
where $\mu$ is the angular momentum.

The order parameter $\Delta(\m{r})$ is the spatially variable which has obvious asymptotic behavior:
\begin{equation}\label{Deltasympt}
\Delta(r)=\cases{0 &if $r \to 0$\\
    \Delta_{0} &if $r \to \infty$\\}
\end{equation}
The limiting case $\varkappa \to \infty$ corresponds to the point-like vortex. As we mostly interested in localized states near the Ab-vortex core where $\Delta$ is suppressed the order parameter $\Delta$-term in \eref{ham0} can be replaced by the proper boundary conditions at $r\to 0$. The inner structure of the vortex is encoded into the parameters of the boundary conditions. In particular in such an approach the divergence of the slope $d\Delta /dr$ at $r\to 0$ can be treated correctly via introducing the parameter $\xi_{1} \left.\,d\Delta /dr\right|_{r=0}$. It controls the specific boundary conditions.

Note that without $\Delta$-potential the Hamiltonian \eref{ham0} coincides with that used for the description of AB effect  where the localization of the magnetic flux also takes place \cite{abeffect_physrev1959}.
The elementary excitation with $\varepsilon \ll \Delta_{\infty}$ are localized near the Ab-vortex line and as was noted in \cite{sc_carollidegennes_physlet1964} play a major role in transport and relaxation phenomena at low temperatures.

It is clear that the presence of a defect influence the spectrum of these excitations.

The applied magnetic field $\sim H_{c_1}$ and therefore the vector-potential is of order $A\sim H_{c_1}\xi$. Therefore with account of
\begin{equation}
\frac{c\hbar}{2e\xi}\nabla\theta\sim H_{c_2}\,,
\end{equation}
and $\frac{H_{c_1}}{H_{c_2}}\sim \frac{1}{\varkappa}<<1$ the applied magnetic field for the excitations of interest
can be omitted:
\begin{equation}\label{ham2}
\hat{H}=\hat{H}_{AB}+\sigma_{x}\Delta\,,
\end{equation}
where
\begin{equation}\label{ham_ab}
  \hat{H}_{AB} = \sigma_{z}\left\{\frac{1}{2m}\left(\,\hat{\mathbf{p}}+
\sigma_{z}\frac{e}{c}\mathbf{A}_{eff} \,\right)^{2}-E_{F}\right\}
\end{equation}
is the part of the Hamiltonian. It incorporates the action of the magnetic field due to the localized vortex
\begin{equation}
 A_{eff}\sim H_{c_2}\xi\,.
\end{equation}

If we take into account point-like character of the magnetic flux distribution:  \[\mathbf{A_{eff}}(r)=\frac{\Phi}{2\pi\,r}\mathbf{e}_{\phi}\,,\]
and use the standard separation of radial and angular variables we obtain the Aharonov-Bohm Hamiltonian (ABH). Its radial part corresponds to that of the BdGH with $\alpha=1/2$ \eref{radham} in the form
\begin{equation}\label{ABham}
\hat{H}_{AB}=\sigma_{z}\frac{\hbar^{2}}{2m}\Bigl\{-\frac{d^{2}}{dr^{2}}-\frac{1}{r}\frac{d}{dr}
+\left(\mu-\alpha\sigma_{z}\right)^{2}\frac{1}{r^{2}}-E_{F}\Bigr\}
\end{equation}
where $\alpha=\{\frac{\Phi}{\Phi_{0}}\}$ is the fractional part of the magnetic flux $\Phi$ in the vortex core and $\Phi_{0}$ is the magnetic flux quantum.

As is known \cite{sc_vortexcorekramerpesch_zphys1974} in the quantum limit the size of the vortex core $\xi_{1}\ll\xi$ then the magnetic flux which localized in the vortex core is defined as $\Phi=\pi\xi_{1}^{2}H_{c_{2}}$.
It is easy to show that $\alpha$ has the form:
\begin{equation}\label{alph}
\alpha=\frac{\Phi}{\Phi_{0}}=
\frac{1}{2}\left(\frac{\xi_{1}}{\xi}\right)^{2}\,,
\end{equation}
where
\[\xi=\sqrt{\frac{\hbar}{m\,\omega_{H_{c_{2}}}}}\,.\]
is Ginzburg-Landau (GL) coherence length.

Also
\[\omega_{H_{c_{2}}}=\frac{2\,e\,H_{c_{2}}}{m\,c}\]
is the cyclotron frequency.

Based on the fact that in the quantum limit $\xi_{1}\sim\,\frac{1}{k_{F}}$ \cite{sc_vortexcorekramerpesch_zphys1974} we obtain
\begin{equation}\label{alph2}
\alpha\simeq\frac{\hbar\,\omega_{H_{c_{2}}}}{4\,E_{F}}
\end{equation}
This suggests that energy of bound state $\hbar\,\omega_{H_{c_{2}}}$ (more detail it will be discussed in section \ref{sec_bound}) having regard to \eref{alph} is proportional to $\left(\frac{\xi_{1}}{\xi}\right)^{2}E_{F}$. Thus the equation for the bound states \eref{ABham}, then it is natural to measure the energy in units of $\left(\frac{\xi_{1}}{\xi}\right)^{2}E_{F}$.

If one takes into account that \[\frac{1}{k_{F}}\,\left.\frac{d\,\Delta}{d\,r}\right|_{r=0}
\simeq\frac{\Delta_{0}}{k_{F}\xi_{0}}
\simeq\frac{\hbar\,\omega_{H_{c_{2}}}}{\pi}\]
then
\begin{equation}\label{alph3}
 \alpha\simeq\frac{\pi}{4\,k_{F}}\,\left.\frac{d\,\Delta/E_{F}}{d\,r}\right|_{r=0}
\end{equation}
where
\[\xi_{0}=\frac{\hbar\,\upsilon_{F}}{\pi\Delta_{0}}\sim \xi\]
is the BCS coherence length, $\Delta_{0}$ the gap at $T=0$ and $\upsilon_{F}$ is the Fermi wave velocity. According to the work \cite{sc_vortexcorelowexcit_prl1998}  for high-temperature superconductors in which $1<k_{F}\xi_{0}<5$ in the quantum limit the value $\xi_{1}/\xi_{0}$ lies in the interval $0.3\div0.6$. Therefore $\alpha\ll\frac{1}{2}$ that leads to the fact that the magnetic flux $\Phi$ in the vortex core is very small compared with $\Phi_{0}$. So for example in the quantum limit for the $YBCO$ ($k_{F}\xi_{0}\sim4$ with $T_{c}=90\,K$ \cite{sc_tunnelingspect_prl1995}), $\xi_{1}/\xi_{0}\sim 0.3$ \cite{sc_vortexcorelowexcit_prl1998} that corresponds to $\alpha\sim 0.05$.

The Hamiltonian \eref{ABham} can be treated by
the zero-range potential method \cite{book_bazeldovichperelomov_en} by specifying the correct boundary conditions which provide self-conjugacy of the Hamiltonian \eref{ABham}. In \cite{abeffect_deltainterct_jmathphys1998} all possible boundary conditions for the Hamiltonian \eref{ham_ab} as well as its spectrum were obtained. The general result is that depending on the type of the boundary conditions there can be no more than two bound states in the singular point where the magnetic flux is localized.
Although AB effect is interesting in itself it can not provide clear physical interpretation for all types of the boundary conditions because the only output is the interference shift.

We show that the existence of the bounded states of the elementary excitations in the the Ab-vortex and the pinning defects can be described within such approach.
We search for the interpretation of these nonstandard boundary conditions found in \cite{abeffect_deltainterct_jmathphys1998} in terms of the vortex pinning because of the close relation between these problems stated above in the considered limit.

\section{Two types of bound states near the vortex core}\label{sec_bound}
In \cite{sc_carollidegennes_physlet1964} the existence of the low lying branch of bounded states which is emerged from the vacuum state of Cooper pairs was proved.
Because of the magnetic nature of the vortex the low lying bound states can be thought of as Landau levels in the effective magnetic field $H_{eff}\approx H_{c_2}$ with the spectrum:
\begin{equation}\label{cdgennes_branch}
E_{\mu} = \hbar\,\omega_{H_{eff}} \,\mu \,,
\end{equation}
where
\begin{equation}\label{omega_eff}
  \omega_{H_{eff}} = \omega_{L} + \frac{1}{k_{F}}\,\left.
  \frac{d\, \Delta }{d\, r}\right|_{r=0}\,.
\end{equation}
In extreme type-II superconductors the first term can be omitted in comparison with the second term which is much greater \cite{sc_vortexcorespectr_physb2002}. Note that the result \eref{omega_eff} does not depend on the specific form of $\Delta(r)$ but only on the limiting value of the radial derivative. This is quite natural as the bound states are strongly localized. Therefore in order to get the ground state for this branch we can omit the $\Delta$-potential term in \eref{ham0}. Then the proper boundary conditions which correspond to the localization of the magnetic flux and in particular the singular behavior of $\frac{d\, \Delta }{d\, r}$ Hamiltonian \eref{ham_ab} should be used.

According to the theory of self-conjugate extensions
for the Hamiltonian \eref{ham_ab} \cite{abeffect_deltainterct_jmathphys1998} there is 4-parametric
set of boundary conditions for each value of $\alpha$. They are represented by the unitary matrix:
\begin{equation}\label{u2}
 \eqalign{U = e^{i\,\omega}
\left(
\begin{array}{cc}
q\,e^{i\,\varphi_1} & \sqrt{1-q^2}e^{i\,\varphi_2}\\
   -\sqrt{1-q^2} e^{-i\,\varphi_2} & q\,e^{-i\,\varphi_1}
\end{array}
\right)\cr
0\le q \le 1, \quad \varphi_i \in \mathbb{R}, \quad i=1,2}\,.
\end{equation}
Note that for the Hamiltonian \eref{ham_ab} $\alpha $ stands for the value of the localized flux. In the case of the Ab-vortex this means that this is the fraction of the flux quantum in the vortex core.

Regardless of the fact that Hamiltonians \eref{ABham} and \eref{ham_ab} coincide, the boundary conditions which determine their spectrum are related to different physical situations. Depending on the type of the boundary condition there can be one, two or none bound states \cite{abeffect_deltainterct_jmathphys1998}.
The case of the absence of the bound states corresponds to the
AB effect itself ($q=1,~\omega=0,~a=\pi,~\alpha\in (0,1)$).

The result of \cite{sc_carollidegennes_physlet1964} allows one to state that in the case of localized Ab-vortex at least one bound state should exist. The bound state which may occur in the AB effect due to the point-like interaction with the material of the solenoid is different in physical nature from that which appear as the bound state of the quasi-particle excitation near the vortex core. These states should be different with respect to their sensitivity to the change of the magnetic flux $\alpha$. We consider this case first.

\subsection{Bound states in the vicinity of the vortex core}
The existence of the only bound state can be related to 1) the AB plus the defect which leads to the localization by its point-like potential $(q<1,~\omega=0,~a=\pi,~\alpha\in (0,1))$; 2) the localized Ab-vortex ($0\leq\,q\leq1,~\omega=0,~a=\pi/2,~\alpha\in (0,1/2)$) where the lowest bound state is due to localization of the excitation near the core surrounded by the Cooper pairs. It is possible to distinguish between these situations considering the dependence of the bound state on the change in magnetic flux ratio $\alpha$. In case of AB bound state which is caused mainly by the defect the energy of this state does not depend essentially on $\alpha$. In contrast to this the bound state for Ab-vortex depends on the flux ratio much stronger. To confirm this statement and to connect the case of the only bound state for the Hamiltonian \eref{ham_ab}  with the result \eref{omega_eff} let us consider the derivative $\frac{\partial\,\varepsilon_{1}}{\partial\,\alpha}$ which is the measure of the ``magnetic sensitivity`` of the corresponding energy level. Note, that both parameters $\alpha$ and $\mu$ enter the Hamiltonian by the same way. Therefore  the value $\frac{\partial\,\varepsilon_{1}}{\partial\,\alpha}$ for the Hamiltonian \eref{ham_ab} can be compared in the quantum limit with the value
\begin{equation}\label{hw}
\left(\frac{\xi}{\xi_{1}}\right)^{2}\frac{d\,E_{\mu}/E_{F}}{d\mu}
\simeq\left(\frac{\xi}{\xi_{1}}\right)^{2}\left[\hbar\omega_{L}/E_{F}+\left(\frac{\Delta_{0}}{E_{F}}\right)^{2}\right]\approx2
\end{equation}
for the BdGH. \Eref{hw} also can be written as $\left(\frac{\xi}{\xi_{1}}\right)^{2}\frac{\hbar\,\omega_{H_{c_{2}}}}{E_{F}}\approx2$ and using the definition of $\alpha$ in \eref{alph} we can obtain that $\alpha\simeq\frac{\hbar\,\omega_{H_{c_{2}}}}{4\,E_{F}}$. This is consistent with \eref{alph2}.

\begin{figure}[th]
\begin{multicols}{2}
\includegraphics[scale=0.7]{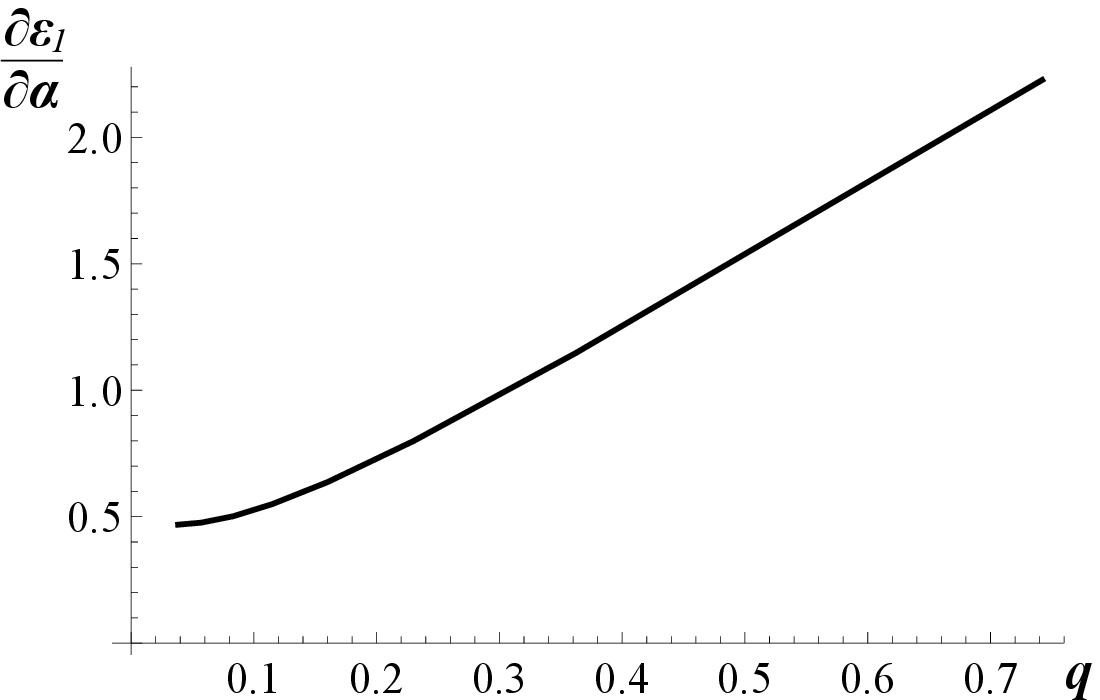}
\caption{The dependence of $\left.\frac{\partial\varepsilon_{1}}{\partial\alpha}\right|_{\Delta_{0}}$ on $q$.}
\label{e_q}
\includegraphics[scale=0.7]{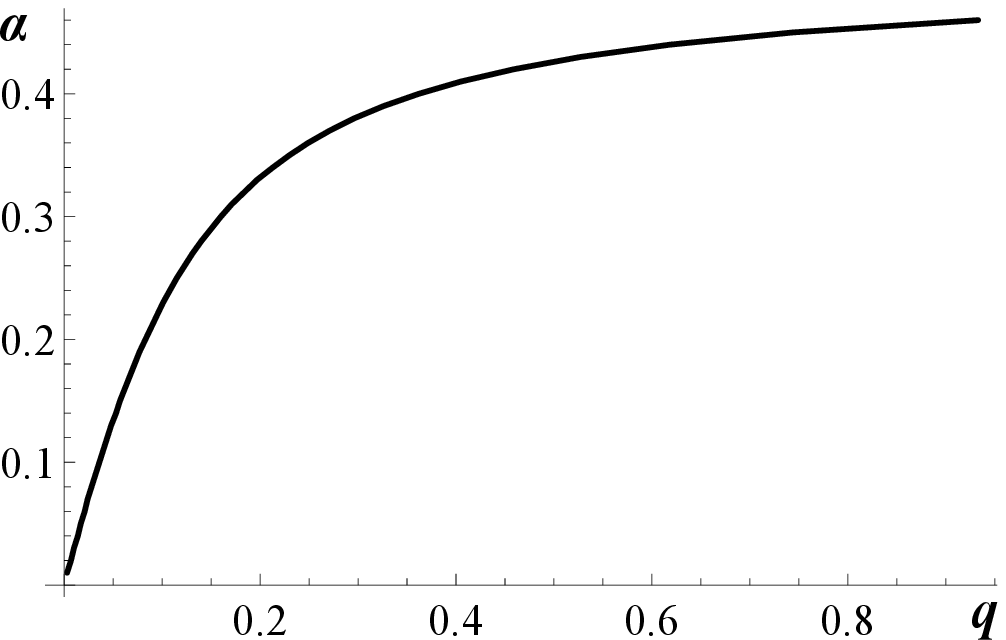}
\caption{The dependence of \mbox{$\alpha$ on $q$.}}
\label{a_q}
\end{multicols}
\end{figure}

To obtain the dependence $q(\alpha,\eta)$ we use the condition
\begin{equation}\label{e1}
-\varepsilon_{1}\left(\,\alpha , q \,\right) = \eta
\end{equation}
to fix the energy gap. Here
\[\eta=\left(\frac{\xi}{\xi_{1}}\right)^{2}\left[\hbar\omega_{L}/E_{F}+\left(\frac{\Delta_{0}}{E_{F}}\right)^{2}\right]\]
in accordance with \eref{hw}.
In accordance with said above the value of the derivative $\left.\frac{\partial\varepsilon_{1}}{\partial\alpha}
\right|_{\Delta_{0}}$ as:
\begin{equation}\label{der}
\left.\frac{\partial\varepsilon_{1}}{\partial\alpha}
\right|_{\Delta_{0}}
=-\left.\frac{\partial\varepsilon_{1}}{\partial\,q}
\right|_{\Delta_{0},\alpha(q)}\frac{dq}{d\alpha}
\end{equation}
The dependence of $\left.\frac{\partial\varepsilon_{1}}
{\partial\alpha}\right|_{\Delta_{0}}$ and $\alpha$
on the parameter $q$ is shown in figure~\ref{e_q} and figure~\ref{a_q} are respectively. Note that the value
 $\frac{\partial\varepsilon_{1}}{\partial\alpha}$ increases  linearly with
 the parameter $q$ and the value $\alpha$ is also increases linearly at low $q$. According to \eref{alph3}--\eref{omega_eff} the same behavior
 $\frac{d\,E_{\mu}}{d\mu}$ and $\alpha$ on $\frac{d\Delta }{dr}$ takes place. It allows to state that the parameter $q$ controls the limiting slope parameter $\xi_{1}\,\frac{d\Delta }{dr}$ at $\varkappa \to \infty$.

It should be noted that the range of values $q(\alpha,\eta)$ depends essentially on the parameter $\eta$. There is the minimum value of $\eta$, when $q$ runs over the interval (0,1) if $0 \le\alpha \le 1/2$. This minimum value is $\eta^{(min)} = \frac{\pi}{2\,W\left[\frac{\pi}{2}\right]}\approx2.11\,,$ where W is Lambert function. This result is in correspondence with the following facts. The first one is that the existence of half-quantum vortices bounds the value of $\alpha$ by $1/2$. The second is that in the quantum limit the value $\eta$ is of order of unit and can not be too small. In the opposite case $\left(\frac{T}{T_{c}}\sim1\right)$ when $\eta\ll1$  there is no dependence between $q$ and $\alpha$, that leads to $\left.\frac{\partial\varepsilon_{1}}{\partial\alpha}\right|_{\Delta_{0}}=0$.

The choice of the parameter $a$ influence the bifurcation between one and two bound states, which occurs with the variation of the flux parameter $\alpha$. Thus in the case of one bound state, the value of parameter $a$ is associated with the maximum value of the parameter $\alpha$ for this bound state. This corresponds to the maximum value of the magnetic flux in the Ab-vortex core. The total flux is distributed between the core of size $\xi_1$ and the outer region where the magnetic field decay exponentially. As was shown \cite{sc_halflux_europhys1996} in epitaxial films of $Bi_{2}Sr_{2}CaCu_{2}O_{8+\delta}$, at the meeting point of a tricrystal substrate of $SrTiO_{3}$ appears half-quantum vortex (HQV). This effect is associated with d-wave nature of the superconductor and together with photoemission results, proves that the in-plane order parameter for this high-Tc cuprate superconductor closely follows $d_{x^{2}-y^{2}}$ symmetry. The existence of the half-quantum vortices allows to suggests that within the vortex core should be no more than half of magnetic flux quantum. Such a restriction on the fraction of the magnetic flux in the vortex core corresponds to the choice of $a=\frac{\pi}{2}$.

\subsection{Bound states generated by the pinning center}
\begin{figure}[th]
\begin{center}
  \includegraphics[scale=1]{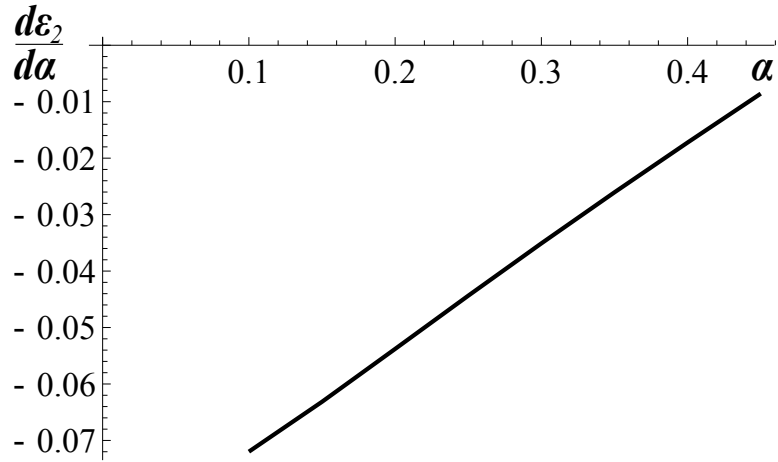}\\
  \caption{The dependence of $\left.\frac{d\varepsilon_{2}}{d\alpha}\right|_{{\Delta}_{\infty}}$ on $\alpha$.}\label{e_alpha}
\end{center}
\end{figure}

The low lying bound states considered above appear due to the interaction with surrounding Cooper pairs which prohibit the propagation of the excitations beyond the core. If the Ab-vortex is pinned by the defect one can expect the appearance of the new bound state. From the physical point of view such bound states may appear as the result of the resonant scattering from the defect from one side and the Cooper pairs in the bulk of the superconductor from the other side. In \cite{sc_vortexpinning_prb2009} it was shown that indeed the additional branch of the bound states appears. If the results in \cite{sc_vortexpinning_prb2009} rewritten in terms of $\mu$, then we can show that this  branch is less ''sensitive'' to magnetic field, which by its nature is different from the bound states  discussed above. This difference is concluded in the behavior of the derivative $\frac{d\,E_{\mu}/E_{F}}{d\mu}$ , which is less than zero. To compare these results with our consider the case when $\omega > \frac{\pi}{2}$ with fixed $a=\pi/2$ which leads to appearance of two roots. The presence of the second bound state is connected with the appearance of additional branch associated with the resonant scattering from the insulating defect. To confirm this statement let us consider the derivative $\left.\frac{d\varepsilon_{2}}
{d\alpha}\right|_{\Delta_{0}}$ which is the measure of the ''magnetic sensitivity'' of the corresponding energy level. It is easy to see that this derivative is not zero in contrast to the previous case. The dependence of $\left.\frac{d\varepsilon_{2}}
{d\alpha}\right|_{\Delta_{0}}$ on the parameter $\alpha$ is shown in figure~\ref{e_alpha}, where shows that $\frac{d\varepsilon_{2}}{d\alpha}$ decreases linearly with the parameter $\alpha$ that is consistent with  behavior $\frac{d\,E_{\mu}/E_{F}}{d\,\mu}$.

\section*{Conclusion}
In the paper it is shown that the consideration of the low lying bound states localized in the vicinity of the vortex core can be investigated with the help of the formalism \cite{abeffect_deltainterct_jmathphys1998} developed for ABH. We show that the nonstandard boundary conditions constructed in \cite{abeffect_deltainterct_jmathphys1998} can be interpreted in terms of the localization of the excitations in the vortex core. It is shown that at least one bound state exists in the vortex core even in the quantum limit when $\xi_1\sim k^{-1}_F$. Within the proposed approach the case of the divergent slope of the order parameter can be treated correctly. Another possibility for the localization can be realized for the pinned vortex due to the resonant scattering of the excitation between the defect and the surrounding Cooper pairs.

\appendix
\section{Characteristics of pinning}
Here we give the variational estimates for the characteristics of pinning. The result \eref{cdgennes_branch} allows to consider the lowest bound states in the core as the Landau levels in the effective magnetic field of order $H_{c_{2}}$ corresponding to the localized flux. The effect of pinning can be considered as the additional localization due to the defect. This gives the ground to choose the trial wave function of the ground state as the superposition of bound states for these potentials correspondingly.

\begin{figure}[th]
\begin{center}
 \includegraphics[scale=1]{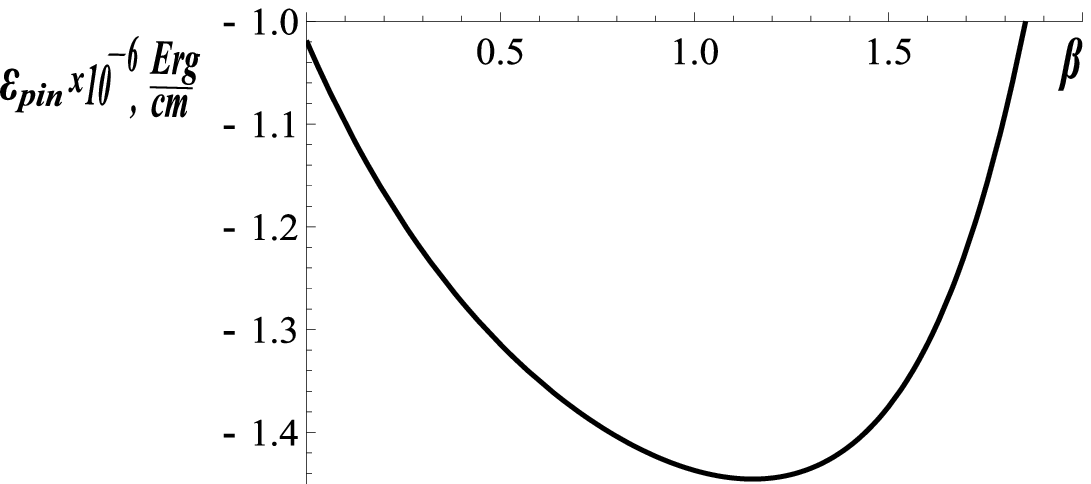}\\
  \caption{The dependence of $\varepsilon_{pin}$ on the variational parameter $\beta$.}\label{fig_epinbeta}
\end{center}
\end{figure}

We choose the trial wave function of the ground state ($m=0$) in the form:
\begin{equation}
\Psi=\Psi_{Land}\sin\beta +\Psi_{bound}\cos\beta
\end{equation}
where
\begin{equation}
\Psi_{Land}=\frac{\exp(-\frac{\rho^{2}}{4a^{2}})}{a}; \qquad \Psi_{bound}=J_{0}(\rho)
\end{equation}
and $a$ cyclotron radius
\[a=\xi=\sqrt{\frac{\Phi_{0}}{2\pi H_{c_2}}}\]
are the corresponding wave functions representing the ground states for corresponding interaction. Pinning energy per unit length calculated based on a variational method for $YBaCuO$ ($\xi_{0}\approx12$~{\AA}, $\lambda_{0}\approx1000$~{\AA} \cite{sc_vortexpincavity_variational_lowtemp2002}). The result for the energy as the function of the variational parameter is shown in figure~\ref{fig_epinbeta}.

The corresponding minimum of the energy is:
\begin{equation}\label{energy}
\varepsilon_{pin}=-1.45\cdot 10^{-6}Erg/cm
\end{equation}
The force of the pinning per unit length is calculated as:
\begin{equation}\label{force}
  f_{pin} = -\left.
  \frac{\partial\, \varepsilon_{pin}}{\partial\, R}\right|_{R = \xi}
\end{equation} and the corresponding critical supercurrent $j=\frac{2e}{h}f_{pin}$ at the used parameters have values \[f_{pin}\approx 18~\frac{Dyn}{cm}\,,\quad j\approx 9 \cdot 10^{8}\frac{A}{cm^2}\]

Using such a simple model the temperature dependence of  $\varepsilon_{pin}$ and $j$ can be obtained. Indeed, according to the equations of \eref{cdgennes_branch}, \eref{force} and  the definition of $j$ for the quantities $\varepsilon_{pin}$ and $j$ the following relations are valid:
\begin{equation}
    \varepsilon_{pin}\propto H_{c_{2}}; \quad
    j_c \propto H_{c_{2}}^{\frac{3}{2}}\,.
\end{equation}
Next, we use the fact that $H_{c_{2}}\sim\tau$, where \mbox{$\tau=1-(T/T_{c})^{2}$} and then we easily find that
\begin{equation}
    \varepsilon_{pin}\propto\tau; \quad
    j_c \propto \tau^{\frac{3}{2}}\,.
\end{equation}

Note that such temperature dependence of $j$ coincides with the result of \cite{sc_vortexpincavity_variational_lowtemp2002}. The calculations of $\varepsilon_{pin}$   performed for $YBaCuO$ in \cite{sc_vortexpincavity_variational_lowtemp2002} give $\varepsilon_{pin}\sim-10^{-6}Erg/cm$. This is consistent with the estimates \eref{energy} obtained above.

\ack
The authors thanks to Prof. Vadim Adamyan for clarifying discussions of the results.


\end{document}